\documentclass[journal,a4paper]{IEEEtran}

\usepackage{cite}
\usepackage{amsmath,amssymb,amsfonts, mathrsfs}
\usepackage{algorithmic}
\usepackage{graphicx}
\usepackage{xcolor}
\usepackage{multirow}
\usepackage{mathtools}

\usepackage[caption=false,font=normalsize,labelfont=sf,textfont=sf]{subfig}

\usepackage{url}
\begin{document}

\title{On the Outage Capacity of the\\Massive MIMO Diversity Channel}

\author{Marco Martal\`o and Riccardo Raheli
\thanks{M.~Martal\`o is with the Department of Electrical and Electronic Engineering, University of Cagliari, Italy. R.~Raheli is with the Department of Engineering and Architecture, Information Engineering Unit, University of Parma, Italy.
E-mail: marco.martalo@unica.it, riccardo.raheli@unipr.it.}
}


\maketitle

\begin{abstract}
We consider the massive Multiple Input Multiple Output (MIMO) diversity channel affected by independent and identically distributed Rayleigh fading, with linear processing at both transmitter and receiver sides, and analyze the outage capacity for large number of antennas. We first discuss the classical Single Input Multiple Output (SIMO) diversity channel that uses Maximal Ratio Combining (MRC) or Selection Combining (SC). For MRC, a numerical computation and a Gaussian Approximation (GA) are considered, whereas for SC an exact evaluation is possible. The analysis is then straightforwardly extended to the Multiple Input Single Output (MISO) system that uses Maximal Ratio Transmission (MRT) or transmit antenna selection. The general Multiple Input Multiple Output (MIMO) system that pursues full diversity is finally considered, with both optimal linear processing and simple antenna selection at both transmitter and receiver. If the number of antennas is sufficiently large on at least one side, the outage capacity of each considered diversity channel approaches that of a suitable reference Additive White Gaussian Noise (AWGN) channel with properly defined Signal-to-Noise Ratio (SNR), which provides a performance benchmark. This conclusion is valid for large but realistic number of antennas compatible with the assumption of independent fading.
\end{abstract}

\begin{IEEEkeywords}
Outage capacity, Diversity channel, Multiple Input Multiple Output (MIMO), Massive MIMO, Maximal Ratio Transmission (MRT), Maximal Ratio Combining (MRC), Selection Transmission and Combining (STC), Rayleigh fading.
\end{IEEEkeywords}

\IEEEpeerreviewmaketitle

\section{Introduction}
\IEEEPARstart{P}{erformance} of wireless communication systems may be strongly limited by fading effects. Such harmful effects can be, however, counteracted by the principle of diversity, which can be exploited in time, frequency or space~\cite{TsVi12}. In particular, space diversity can be obtained by means of sufficiently spaced antennas at the transmitter or receiver.

Space diversity is a classic technique that is acquiring further importance as a key enabler of modern wireless technology. Diversity techniques can be exploited for improving the performance of mobile Internet of Things (IoT) systems in faded environments~\cite{BaShKa20}. Massive arrays of antennas are considered in modern 5G communications and beyond, see, e.g.,~\cite{ZhBjMaNgYaLo20} and references therein. Moreover, diversity is considered to fulfill the requirements of ultra-high reliability within a stringent latency constraint~\cite{KhViYu21}. However, large diversity orders may be necessary for acceptable performance, as shown in~\cite{PoStNiCaAnTrBa19,JoWaErHe15} for Rayleigh fading.

Inspired by these recent contributions, in this paper we consider the massive Multiple Input Multiple Output (MIMO) diversity channel, in which a large number of transmit and receive antennas are affected by independent and identically distributed (i.i.d.) Rayleigh fading. This model is, for instance, of interest in distributed massive MIMO systems, in which macrodiversity is achieved by means of several multi-antenna base stations~\cite{DiQuLiCh21,LiLvZhWaWaYo20}. MIMO diversity is of particular interest in applications operating at small Signal-to-Noise Ratio (SNR), such as in low-rate (e.g., IoT) systems, where no multiplexing gain can be achieved by multiple antennas~\cite{TsVi12}. Another scenario where the massive MIMO diversity channel may arise is the so-called doubly massive MIMO, where both transmitter and receiver are equipped with large antenna arrays~\cite{BuDa18}.

In this paper, we analyze the outage capacity of the MIMO diversity channel for large number of antennas on at least one side. The outage capacity is relevant in slow fading channels, in which the ergodic capacity vanishes~\cite{TsVi12}, as one may not be able to transmit codewords with arbitrarily small error probability and reasonably large blocklength, due to the non-zero probability that the channel is in deep fade. We provide a benchmark to the outage capacity of the massive MIMO diversity channel in terms of the capacity of an Additive White Gaussian Noise (AWGN) channel operated at a properly defined Signal-to-Noise Ratio (SNR). This benchmark is obtained for asymptotically high diversity order and proves useful to characterize the massive MIMO diversity channel for large but realistic finite diversity orders of practical interest.

We start from the Single Input Multiple Output (SIMO) system that uses Maximal Ratio Combining (MRC) or Selection Combining (SC). A closed-form expression of the outage capacity of the MRC diversity channel is not available, but numerical evaluation is provided. A Gaussian Approximation (GA) valid for large number of antennas is also discussed and compared. On the other hand, an exact computation of the outage capacity is possible for the SC diversity channel. This analysis can be straightforwardly extended to the Multiple Input Single Output (MISO) system with either Maximal Ratio Transmission (MRT)~\cite{Lo99} or transmit antenna selection, here referred to as Selection Transmission (ST).

The general case of MIMO systems pursuing full diversity is finally analyzed. We assume optimal linear processing at both transmitter and receiver sides in order to achieve the maximum diversity order given by the product of the number of transmit and receive antennas. We also consider the simple antenna selection scheme at both transmitter and receiver, here referred to as Selection Transmission and Combining (STC). 

Our results show that, if the number of antennas is sufficiently large, the outage capacity of the diversity channel closely approaches the benchmark provided by the AWGN channel with properly defined SNR but for a gap, under operational condition compatible with the independent fading assumption. Bounds and asymptotic results are also provided for the massive MIMO channel.

The structure of this paper is the following. In Section~\ref{sec:system-model}, we review the MIMO, SIMO, and MISO diversity channels. In Section~\ref{sec:outage_capacity}, we derive the outage capacity benchmarks for the considered scenarios. Numerical results are discussed in Section~\ref{sec:results}. Finally, concluding remarks are given in Section~\ref{sec:concl}.

\section{MIMO Diversity Channel} \label{sec:system-model}
\begin{figure}
\centering
\includegraphics[width=0.48\textwidth]{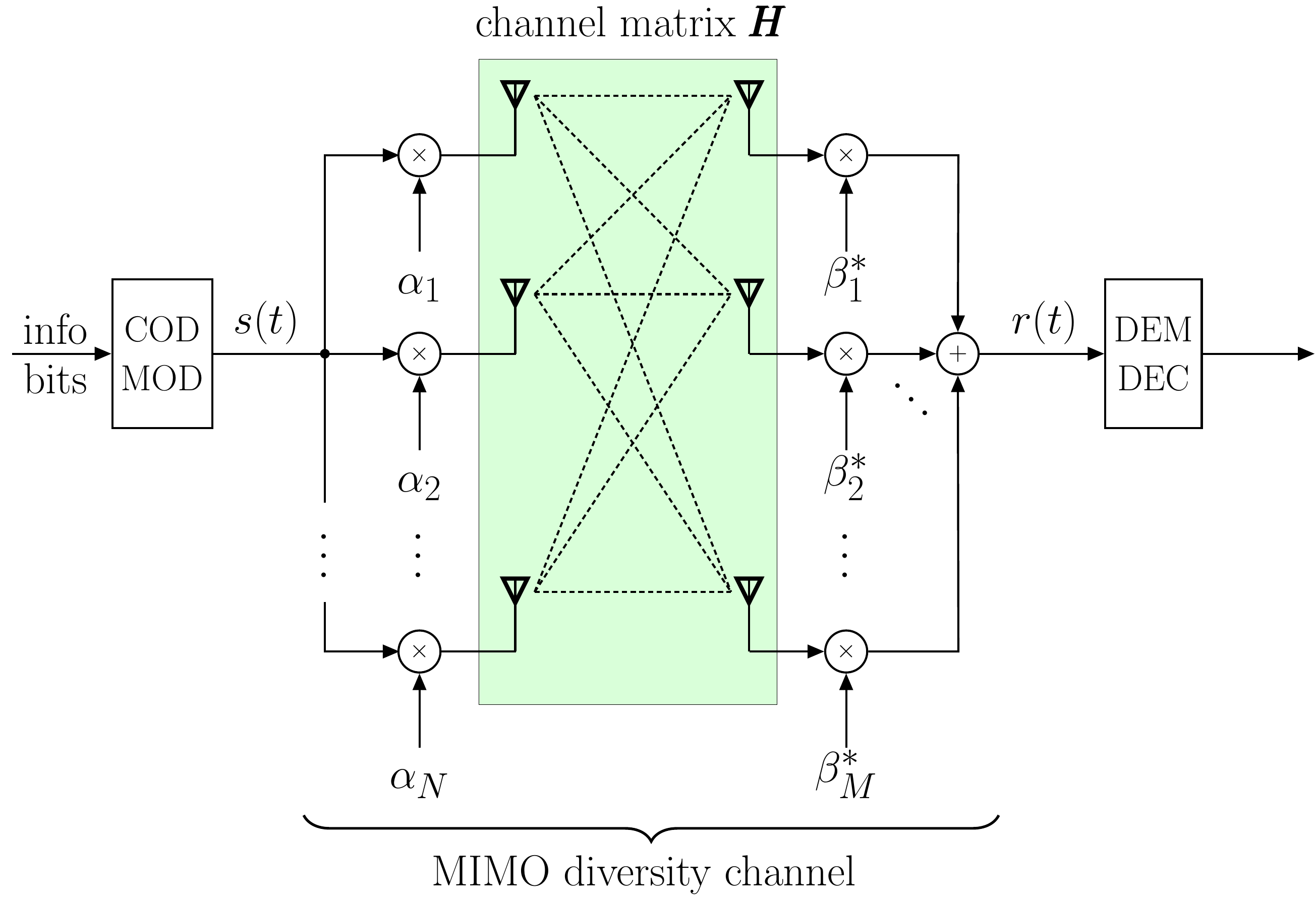}
\caption{MIMO diversity channel with $N$ transmit and $M$ receive antennas. Transmit beamforming and receive linear processing are considered.}
\label{fig:scenario}
\end{figure}

Consider the MIMO diversity channel depicted in Fig.~\ref{fig:scenario}. 
The encoded and modulated transmitted signal $s(t)$ is precoded by the beamforming vector $\pmb{\alpha}=[\alpha_1,\alpha_2,\ldots,\alpha_N]^T$ and transmitted through $N$ antennas, where the symbol $T$ denotes the transpose operator. The channel matrix $\pmb{H}$ has size $M\times N$ and independent frequency-flat slow fading gains $\{h_{ij}\}$. At the receiver, signals received by $M$ antennas are linearly combined by the receive vector $\pmb{\beta}^*=[\beta_1,\beta_2,\ldots,\beta_M]^\dagger$, where $\dagger$ denotes the Hermitian operator. The receive antennas are also affected by the AWGN vector $\pmb{w}(t)=[w_1(t),w_2(t),\ldots,w_M(t)]^T$. The combined signal is used for demodulation and decoding.

The fading gains are assumed i.i.d. and follow the Rayleigh model. This independence assumption can be justified for massive antenna arrays of practical interest. As an illustrative example, assume that antenna elements are spaced by half wavelength to observe i.i.d. fading gains and consider a carrier frequency $f_{\rm c}=5$~GHz, i.e., a wavelength $\lambda=c/f_{\rm c}=6$~cm, where $c=3\cdot10^8$~m/s is the propagation speed. Then, antenna elements must be spaced at least by 3 cm for independency. Therefore, a massive square array of $100\times100=10^4$ antenna elements has dimension of approximately $3\times3$~m, which can be considered realistic for cellular base stations.

The received signal can be expressed as
$$
r(t) = \pmb{\beta}^\dagger \pmb{H} \pmb{\alpha} x(t) + n(t)
$$
where $n(t)=\pmb{\beta}^\dagger\pmb{w}(t)$ is the noise signal at the output of the receiver linear processor.

Among various linear processing schemes in the literature, we focus on two possible approaches. First, we consider optimal beamforming and receive combining to maximize the combiner SNR, i.e., the SNR at the input of the demodulator. Assuming $||\pmb{\alpha}||=||\pmb{\beta}||=1$ for signal and noise normalization purposes, it is well known that the SNR is maximized if $\pmb{\alpha}$ and $\pmb{\beta}$ are the principal right and left singular vectors of $\pmb{H}$, respectively~\cite{TsVi12}. Under this assumption, it can be verified that the instantaneous SNR at the input of the demodulator can be expressed as~\cite{TsVi12}
\begin{equation} \label{eq:SNR_MIMO}
\gamma_{\rm c}= \rho \sigma^2_{\rm max}
\end{equation}
where $\rho$ is the SNR on a single unit-gain link and $\sigma_{\rm max}$ is the largest singular value of $\pmb{H}$.

Another possible approach, referred to as Selection Transmission and Combining (STC), selects the transmit-receive antenna pair which exhibits the maximum instantaneous SNR. The combiner SNR is therefore
\begin{equation} \label{eq:SNR_STC}
\gamma_{\rm c}=\max_{i,j} \gamma_{ij} 
\end{equation}
where $\gamma_{ij}$ is the SNR on the communication link between the $j$-th transmit antenna, used at full power, and $i$-th receive antenna ($j=1,2,\ldots,N$ and $i=1,2,\ldots,M$). This approach is described by the linear processor with beamforming vector $\pmb{\alpha}$ and receive combiner $\pmb{\beta}$ characterized by just one unitary element, whereas all other elements are zero. By direct extension of the results in~\cite{Br03}, relative to the classical SC in the SIMO case, the average SNR at the output of the combiner has the following closed-form expression for i.i.d. Rayleigh fading:
\begin{equation} \label{eq:avg_SNR_STC}
\overline{\gamma}_{\rm c} = \overline{\gamma}\sum_{\ell=1}^{MN}\frac{1}{\ell} .
\end{equation}

In the next subsections, we discuss the special cases of SIMO and MISO channels.

\subsection{Receive Diversity (SIMO)} \label{subsec:sysmod_SIMO}
In the case of a single-antenna transmitter, $N=1$, the system in Fig.~\ref{fig:scenario} collapses to a standard SIMO system with receive diversity employing $M$ receive antennas. This scenario may be representative of uplink communications between single-antenna users and a multi-antenna base station.

In this case, $\pmb{\alpha}=1$ and the optimal linear combination at the receiver is MRC, i.e., $\pmb{\beta}=\pmb{h}/||\pmb{h}||$, where $\pmb{h}=[h_1,h_2,\ldots,h_M]^T$ is the vector of channel coefficients. It is well known that the combiner SNR is~\cite{TsVi12}
\begin{equation} \label{eq:SNR_MRC}
\gamma_{\rm c} = \sum_{\ell=1}^M \gamma_\ell
\end{equation}
where $\gamma_\ell$ is the SNR on the $\ell$-th receive antenna. Its average value is
$$
\overline{\gamma}_{\rm c} = \sum_{\ell=1}^M \overline{\gamma}_\ell
$$
which reduces, for i.i.d. fading, to $\overline{\gamma}_{\rm c}=M\overline{\gamma}$.

For $N=1$, the STC approach (\ref{eq:SNR_STC}) becomes the well-known SC, i.e., the combiner selects the signal of the diversity branch which exhibits the maximum instantaneous SNR. The combiner SNR is therefore
\begin{equation} \label{eq:SNR_SC}
\gamma_{\rm c}=\max\{\gamma_1,\gamma_2,\ldots,\gamma_M\}
\end{equation}
and the average SNR at the output of the combiner has the following closed-form expression for i.i.d. Rayleigh fading~\cite{Br03}:
\begin{equation} \label{eq:avg_SNR_SC}
\overline{\gamma}_{\rm c} = \overline{\gamma}\sum_{\ell=1}^M\frac{1}{\ell} .
\end{equation}

\subsection{Transmit Diversity (MISO)}\label{subsec:sysmod_MISO}
In the case of a single-antenna receiver, $M=1$, the system in Fig.~\ref{fig:scenario} collapses to a MISO system with transmit diversity employing $N$ transmit antennas. This scenario can be representative of downlink communications between a multi-antenna base station and single-antenna users.

In this case, $\pmb{\beta}=1$ and the optimal linear processor at the transmitter is MRT, i.e., $\pmb{\alpha}=\pmb{h}^*/||\pmb{h}||$. By similar arguments as for MRC, the combiner SNR can be shown to have similar formulation as in (\ref{eq:SNR_MRC}), i.e.,
\begin{equation} \label{eq:SNR_MRT}
\gamma_{\rm c} = \sum_{\ell=1}^N \gamma_\ell
\end{equation}
where $\gamma_\ell$ is the SNR received by the $\ell$-th transmit antenna used at full power~\cite{TsVi12}. The average value of (\ref{eq:SNR_MRT}) is
$$
\overline{\gamma}_{\rm c} = \sum_{\ell=1}^N \overline{\gamma}_\ell
$$
which reduces, for i.i.d. fading, to $\overline{\gamma}_{\rm c}=N\overline{\gamma}$.

For $M=1$, the STC approach (\ref{eq:SNR_STC}) reduces to the well-known ST, which operates as SC, but at the transmitter side. In particular, the combiner SNR is
\begin{equation} \label{eq:SNR_ST}
\gamma_{\rm c}=\max\{\gamma_1,\gamma_2,\ldots,\gamma_N\}
\end{equation}
with average value for i.i.d. Rayleigh fading:
\begin{equation} \label{eq:avg_SNR_ST}
\overline{\gamma}_{\rm c} = \overline{\gamma}\sum_{\ell=1}^N\frac{1}{\ell} .
\end{equation}

\section{Outage Capacity Analysis}\label{sec:outage_capacity}

The outage capacity $C_\varepsilon$ of a slow fading channel is defined as the maximum achievable rate such that the outage probability is less than $\varepsilon$~\cite{TsVi12}.

Let us define the following function
\begin{equation} \label{eq:AWGN_cap}
C(\gamma) = \log_2(1+\gamma)
\end{equation}
which describes the capacity per unit bandwidth of the bandlimited AWGN channel operating at SNR $\gamma$. The outage capacity per unit bandwidth can be expressed as~\cite[Chap.~5]{TsVi12}
\begin{equation} \label{eq:outage_definition}
C_\varepsilon = C\left(\gamma_0\right) = C\left(F^{-1}(\varepsilon)\right) \qquad \textrm{b/s/Hz}
\end{equation}
where $\gamma_0=F^{-1}(\varepsilon)$ is the SNR outage threshold, $F(\cdot)$ is Cumulative Distribution Function (CDF) of the SNR, and $F^{-1}(\cdot)$ is its inverse. This expression is valid for the considered diversity channels in Section~\ref{sec:system-model}, provided $F(\cdot)$ is the CDF of the SNR at the input of the demodulator.

In the following, we shall compare the outage capacity of the diversity channel with that of a reference AWGN channel with suitable SNR. We consider the different scenarios (SIMO, MISO, MIMO) and derive capacity benchmarks for the resulting diversity channels. For simplicity, we start from the standard SIMO case and, then, extend the results to the other cases. To get insights, we also concentrate on the high and low SNR regimes, since the impact of fading may depend very much on the operating regime. In particular, at high SNR the capacity difference between the outage capacity of the investigated diversity channel and the capacity of the relevant reference AWGN one is considered, whereas at low SNR the ratio between these quantities may be more convenient.

\subsection{SIMO} \label{subsec:outage_SIMO}
\subsubsection{Analysis for MRC}
In MRC with i.i.d. branches subject to Rayleigh fading, the SNR (\ref{eq:SNR_MRC}) at the output of the combiner has chi-square distribution with $2M$ degrees of freedom~\cite{TsVi12}. Therefore, the outage probability can be expressed in terms of the CDF of (\ref{eq:SNR_MRC}) as
\begin{equation} \label{eq:CDF_MRC}
\varepsilon = F(\gamma_0) = 1-e^{-\gamma_0/\overline{\gamma}}\sum_{\ell=0}^{M-1} \frac{1}{\ell!}\left(\frac{\gamma_0}{\overline{\gamma}}\right)^\ell .
\end{equation}
Since (\ref{eq:CDF_MRC}) does not admit exact inversion, the computation of the outage capacity $C_\varepsilon^{\rm MRC}$ can be pursued numerically, as it is done in Section~\ref{sec:results}.

However, in order to get insights in the behavior for large $M$, we consider a GA of the random variable (\ref{eq:SNR_MRC}) by the central limit theorem as $\gamma_{\rm c}\sim\mathcal{N}(M\overline{\gamma};M\overline{\gamma}^2)$, where $\mathbb{E}\{\gamma_{\rm c}\}=M\overline{\gamma}$ and ${\rm var}\{\gamma_{\rm c}\}=M\overline{\gamma}^2$, as it can be easily verified. Hence, the outage probability can be approximated as
$$
\varepsilon = F(\gamma_0) \simeq 1 - Q\left(\frac{\gamma_0-M\overline{\gamma}}{\sqrt{M}\overline{\gamma}}\right)
$$
where $Q(x)$ is the tail function of a standard Gaussian random variable. This approximation can be inverted as
\begin{equation} \label{eq:gamma0_MRC}
\gamma_0 = F^{-1}(\varepsilon) \simeq \overline{\gamma}\left[M-\sqrt{M}Q^{-1}(\varepsilon)\right]
\end{equation}
where $Q^{-1}(\cdot)$ is the inverse of $Q(\cdot)$ and the symmetry $Q^{-1}(1-\varepsilon)=-Q^{-1}(\varepsilon)$ has been used. Since the threshold SNR $\gamma_0\geq0$, from (\ref{eq:gamma0_MRC}) the condition $M\geq[Q^{-1}(\varepsilon)]^2$ arises. As shown by the numerical results in Section~\ref{sec:results}, this condition also affects the quality of the approximation, as lower values of $\varepsilon$ require larger values of $M$ for similar accuracy. This leads to the high-order diversity approximation of the outage capacity
$$
C_\varepsilon^{\rm GA}(\overline{\gamma}) = C\left(\overline{\gamma}\left(M-\sqrt{M}Q^{-1}(\varepsilon)\right)\right) .
$$

Considering now a reference AWGN channel at the average branch SNR $\overline{\gamma}$, the resulting capacity difference can be expressed as
\begin{equation} \label{eq:MRC_difference_gamma_GA}
C_\varepsilon^{\rm GA}(\overline{\gamma})-C(\overline{\gamma}) = \log_2\frac{1+\overline{\gamma}\left[M-\sqrt{M}Q^{-1}(\varepsilon)\right]}{1+\overline{\gamma}}
\end{equation}
which grows unboundedly for increasing values of $M$ and any fixed values of $\overline{\gamma}$ and $\varepsilon$. Note that (\ref{eq:MRC_difference_gamma_GA}) can become quite large for practical values  of $M$ such that the i.i.d. fading assumption is still realistic.

By similar arguments, the ratio
\begin{equation} \label{eq:capacity_ratio_gamma_MRC}
\frac{C_\varepsilon^{\rm GA}(\overline{\gamma})}{C(\overline{\gamma})} = \frac{\log_2 \left(1+\overline{\gamma}\left[M-\sqrt{M}Q^{-1}(\varepsilon)\right]\right)}{\log_2 \left(1+\overline{\gamma}\right)}
\end{equation}
grows unboundedly for increasing values of $M$ and fixed values of $\overline{\gamma}$ and $\varepsilon$. 

The growth rate of (\ref{eq:MRC_difference_gamma_GA}) and (\ref{eq:capacity_ratio_gamma_MRC}) could be analyzed. However, a more interesting viewpoint can be obtained excluding the so-called array gain of MRC from the analysis~\cite{TsVi12}. Recall that the array gain is defined as the SNR gain provided by diversity in the absence of fading, i.e., when all the branches are affected by AWGN with deterministic SNR $\overline{\gamma}$. According to (\ref{eq:SNR_MRC}), the SNR in MRC and AWGN is $\gamma_{\rm c}=M\overline{\gamma}$. Since a similar relation holds for the average SNR in i.i.d. fading, $\overline{\gamma}_{\rm c}=M\overline{\gamma}$, we can exclude the array gain from the analysis by considering the capacity difference with respect to the reference AWGN channel with SNR $\overline{\gamma}_{\rm c}$. Using $\overline{\gamma}=\overline{\gamma}_{\rm c}/M$ in (\ref{eq:MRC_difference_gamma_GA}), we have
$$
C_\varepsilon^{\rm GA}(\overline{\gamma}_{\rm c})-C(\overline{\gamma}_{\rm c}) = \log_2\frac{1+\overline{\gamma}_{\rm c}\left[1-\frac{1}{\sqrt{M}}Q^{-1}(\varepsilon)\right]}{1+\overline{\gamma}_{\rm c}} .
$$
For increasing values of $M$, the term $Q^{-1}(\varepsilon)/\sqrt{M}$ tends to 0 for any fixed value of $\varepsilon$. Therefore, for high-order diversity, the outage capacity with MRC approaches from below that of the reference AWGN channel with SNR $\overline{\gamma}_{\rm c}$. Hence, the capacity of this AWGN channel with SNR $\overline{\gamma}_{\rm c}$ provides a benchmark to the outage capacity of the massive SIMO diversity channel.

At high SNR specified by the conditions $\overline{\gamma}_{\rm c}\gg1$ and $\overline{\gamma}_{\rm c}(1-1/\sqrt{M}Q^{-1}(\varepsilon))\gg1$, one can approximate this capacity gap as
\begin{eqnarray*}
C_\varepsilon^{\rm GA}(\overline{\gamma}_{\rm c})-C(\overline{\gamma}_{\rm c}) &\simeq& \log_2 \left[1-\frac{1}{\sqrt{M}}Q^{-1}(\varepsilon)\right]\\ 
&\simeq& -\frac{1}{\sqrt{M}} \frac{Q^{-1}(\varepsilon)}{\ln2}
\end{eqnarray*}
where the second approximation holds because $\ln(1-x)\simeq x$ for $|x|\ll1$. This gap tends to zero from below as $M^{-1/2}$ for increasing values of $M$. Note that the high-SNR condition $\overline{\gamma}_{\rm c}(1-1/\sqrt{M}Q^{-1}(\varepsilon))\gg1$ is well behaved for large $M$, since if it is satisfied for some $M$, it is verified even better for larger values of $M$.

By similar arguments, using $\overline{\gamma}=\overline{\gamma}_{\rm c}/M$ in (\ref{eq:capacity_ratio_gamma_MRC}) the capacity ratio 
$$
\frac{C_\varepsilon^{\rm GA}}{C(\overline{\gamma}_{\rm c})} = \frac{\log_2\left(1+\overline{\gamma}_{\rm c}\left[1-\frac{1}{\sqrt{M}}Q^{-1}(\varepsilon)\right]\right)}{\log_2(1+\overline{\gamma}_{\rm c})}
$$
approaches 1 for increasing values of $M$. Again, this conclusion indicates that $C(\overline{\gamma}_{\rm c})$ describes a benchmark to the outage capacity. At low SNR specified by the conditions $\overline{\gamma}_{\rm c}\ll1$ and $\overline{\gamma}_{\rm c}(1-1/\sqrt{M}Q^{-1}(\varepsilon))\ll1$, the approximate ratio is
\begin{equation} \label{eq:SNR_low_MRC}
\frac{C_\varepsilon^{\rm GA}(\overline{\gamma}_{\rm c})}{C(\overline{\gamma}_{\rm c})} \simeq 1 - \frac{1}{\sqrt{M}} Q^{-1}(\varepsilon) .
\end{equation}
As the low-SNR condition is also well behaved for increasing $M$, we can conclude that this approximation approaches 1 from below as $M^{-1/2}$.

\subsubsection{Analysis for SC}
In SC with i.i.d. branches subject to Rayleigh fading, the outage probability can be expressed as
$$
\varepsilon = F(\gamma_0) = \left(1-e^{-\gamma_0/\overline{\gamma}}\right)^M
$$
which can be inverted as
$$
\gamma_0 = F^{-1}(\varepsilon) = \overline{\gamma}\ln\frac{1}{1-\varepsilon^{1/M}}
$$
where $F(\cdot)$ and $F^{-1}(\cdot)$ are the CDF of the SNR (\ref{eq:SNR_SC}) and its inverse, respectively. The outage capacity is then expressed in terms of (\ref{eq:AWGN_cap}) as
\begin{equation} \label{eq:outage_capacity_SC}
C_\varepsilon^{\rm SC}(\overline{\gamma}) = C\left(\overline{\gamma}\ln\frac{1}{1-\varepsilon^{1/M}}\right) \quad \textrm{b/s/Hz}
\end{equation}
and the difference with respect to that of the reference AWGN channel with SNR $\overline{\gamma}$ is
\begin{equation} \label{eq:SC_difference_gamma}
C_\varepsilon^{\rm SC}(\overline{\gamma})-C(\overline{\gamma}) = \log_2\frac{1+\overline{\gamma}\ln\frac{1}{1-\varepsilon^{1/M}}}{1+\overline{\gamma}}
\end{equation}
which grows unboundedly as $M$ increases for any fixed values of $\overline{\gamma}$ and $\varepsilon$. In fact, using a first-order Taylor series expansion of $1-\varepsilon^x$ about $x\simeq0$ one has
\begin{equation} \label{eq:Taylor}
\lim_{M\rightarrow+\infty}\ln \frac{1}{1-\varepsilon^{1/M}} = \lim_{M\rightarrow+\infty}\ln\left(\frac{M}{-\ln\varepsilon}\right) = +\infty
\end{equation}
since $-\ln\varepsilon\geq0$.

In order to analyze the behaviour of (\ref{eq:SC_difference_gamma}) for large diversity order, one can consider a high-SNR regime, i.e., $\overline{\gamma}\gg1$ and $\overline{\gamma}\ln(1/(1-\varepsilon^{1/M}))\gg1$. Note that, for a given $\varepsilon$, the second high-SNR condition is better verified for increasing $M$. The following approximation, therefore, holds for large $M$
$$
C_\varepsilon^{\rm SC}(\overline{\gamma})-C(\overline{\gamma}) \simeq \log_2\ln\frac{1}{1-\varepsilon^{1/M}}.
$$
This means that the growth is as $\log_2\ln M$, hence significantly slower than MRC.

Unlike MRC, we cannot recognize an array gain in SC due to the fact that in the absence of fading the combiner SNR is $\overline{\gamma}$ for every $M$. However, considering the average SNR (\ref{eq:avg_SNR_SC}), we can identify a gain at the output of the combiner. Using $\overline{\gamma}=\overline{\gamma}_{\rm c}/(\sum_{\ell=1}^M1/\ell)$ in (\ref{eq:outage_capacity_SC}), the outage capacity for SC can be expressed as
$$
C_\varepsilon^{\rm SC}(\overline{\gamma}_{\rm c}) = C\left(1+\overline{\gamma}_{\rm c}\frac{1}{\sum_{\ell=1}^M 1/\ell}\ln\frac{1}{1-\varepsilon^{1/M}}\right) .
$$
Considering the difference with respect to the capacity of the reference AWGN channel with SNR $\overline{\gamma}_{\rm c}$, we obtain
$$
C_\varepsilon^{\rm SC}(\overline{\gamma}_{\rm c})-C(\overline{\gamma}_{\rm c}) = \log_2\frac{1+\overline{\gamma}_{\rm c}\frac{1}{\sum_{\ell=1}^M 1/\ell}\ln\frac{1}{1-\varepsilon^{1/M}}}{1+\overline{\gamma}_{\rm c}} .
$$
As shown below, one has
\begin{equation} \label{eq:limit_SC_quantity}
\lim_{M\rightarrow+\infty} \frac{1}{\sum_{\ell=1}^M 1/\ell}\ln\frac{1}{1-\varepsilon^{1/M}} = 1
\end{equation}
and, therefore, the capacity gap $C_\varepsilon^{\rm SC}(\overline{\gamma}_{\rm c})-C(\overline{\gamma}_{\rm c})$ approaches zero from below for large diversity order, i.e., the outage capacity with SC approaches the benchmark specified by the reference AWGN channel with SNR $\overline{\gamma}_{\rm c}$. To show (\ref{eq:limit_SC_quantity}), the following approximation of the partial sum of the divergent harmonic series can be used
\begin{equation} \label{eq:series_approximation}
\sum_{\ell=1}^M \frac{1}{\ell} \simeq \ln M + k_1 + k_M
\end{equation}
where $k_1\simeq0.57$ is the Euler-Mascheroni constant and $k_M\rightarrow0$ as $M$ increases~\cite[Chap.~5]{Ha10}. Using (\ref{eq:Taylor}) and (\ref{eq:series_approximation}), one has
\begin{eqnarray}
&&\lim_{M\rightarrow+\infty} \frac{1}{\sum_{\ell=1}^M 1/\ell}\ln\frac{1}{1-\varepsilon^{1/M}} \nonumber \\
&&\hspace*{10mm}= \lim_{M\rightarrow+\infty}\frac{1}{\ln M}\ln\left(\frac{M}{-\ln\varepsilon}\right) \nonumber \\
&&\hspace*{10mm}=\lim_{M\rightarrow+\infty} 1 - \frac{\ln(-\ln(\varepsilon))}{\ln M} \label{eq:identity} \\
&&\hspace*{10mm}= 1 . \nonumber
\end{eqnarray}

To analyze the outage capacity for large diversity order, let us consider a sufficiently large SNR, i.e., $\overline{\gamma}_{\rm c}\gg1$ and $\overline{\gamma}_{\rm c}\ln(1/(1-\varepsilon^{1/M})/\sum_{\ell=1}^M (1/\ell))\gg1$, such that the following approximation holds:
\begin{equation} \label{eq:outage_SC_Lfix}
C_\varepsilon^{\rm SC}(\overline{\gamma}_{\rm c})-C(\overline{\gamma}_{\rm c}) \simeq \log_2\left(\frac{1}{\sum_{\ell=1}^M 1/\ell}\ln\frac{1}{1-\varepsilon^{1/M}}\right)
\end{equation}
which, for large $M$, can be approximated using (\ref{eq:identity}) as
$$
C_\varepsilon^{\rm SC}(\overline{\gamma}_{\rm c})-C(\overline{\gamma}_{\rm c}) \simeq \frac{1}{\ln2}\left(-\frac{\ln(-\ln\varepsilon)}{\ln M}\right) .
$$
Using the same Taylor series expansion in (\ref{eq:Taylor}), one obtains that $C_\varepsilon^{\rm SC}(\overline{\gamma}_{\rm c})$ approaches the benchmark $C(\overline{\gamma}_{\rm c})$ from below as $\ln M$. Note that the second high-SNR condition behind (\ref{eq:outage_SC_Lfix}) is also well behaved for increasing $M$ since
$$
\overline{\gamma}_{\rm c}\underbrace{\frac{1}{\sum_{\ell=1}^M 1/\ell}\ln\frac{1}{1-\varepsilon^{1/M}}}_{\rightarrow1 \textrm{ for $M\rightarrow+\infty$}} \gg 1 .
$$

Finally, by similar arguments one can show that the ratio $C_\varepsilon^{\rm SC}(\overline{\gamma}_{\rm c})/C(\overline{\gamma}_{\rm c})$ approaches 1 from below for large values of~$M$. At low SNR, a well behaved approximation is
$$
\frac{C_\varepsilon^{\rm SC}}{C(\overline{\gamma}_{\rm c})} \simeq \frac{1}{\sum_{\ell=1}^M 1/\ell}\ln\frac{1}{1-\varepsilon^{1/M}}
$$
which approaches 1 from below as $(\ln M)^{-1}$.

\subsection{MISO} \label{subsec:outage_MISO}
Since MRT and ST with $N$ branches in MISO systems are equivalent to, respectively, MRC and SC with $N$ branches in SIMO systems as per the results in Sections~\ref{subsec:sysmod_SIMO} and~\ref{subsec:sysmod_MISO}, the analysis in Section~\ref{subsec:outage_SIMO} is also valid here provided $M$ is replaced by $N$.

\subsection{MIMO} \label{subsec:outage_MIMO}
The analysis in Section~\ref{subsec:outage_SIMO} can be extended to massive MIMO scenarios with $N$ transmit and $M$ receive antennas used to pursue full diversity. In particular, we consider the optimal transmit beamforming and receive combining described in Section~\ref{sec:system-model}.

The combiner SNR in (\ref{eq:SNR_MIMO}) can be upper and lower bounded as follows. For any realization of the channel matrix $\pmb{H}$, its largest squared singular value  can be upper and lower bounded by the sum and the average of all the squared singular values, respectively. In other words,
$$
\frac{1}{R_H}\sum_{i=1}^{R_H}\sigma_i^2 \leq \sigma_{\rm max}^2 \leq \sum_{i=1}^{R_H}\sigma_i^2
$$
where $R_H$ is the rank of $\pmb{H}$. Since $R_H\leq\min\{M,N\}$, a further lower bound is obtained as
\begin{equation} \label{eq:bounds_SNR}
\frac{1}{\min\{M,N\}}\sum_{i=1}^{R_H}\sigma_i^2 \leq \sigma_{\rm max}^2 \leq \sum_{i=1}^{R_H}\sigma_i^2 .
\end{equation}
Moreover, it is well known that~\cite[Chap.~3]{PaNaGo03}
$$
\sum_{i=1}^{R_H}\sigma_i^2 = ||\pmb{H}||_{\rm F}^2
$$
where $||\pmb{H}||_{\rm F}$ denotes the Frobenius norm of the channel matrix $\pmb{H}$. Using (\ref{eq:SNR_MIMO}) and (\ref{eq:bounds_SNR}), one finally obtains
\begin{equation} \label{eq:bounds_SNR2}
\underbrace{\frac{\rho}{\min\{M,N\}}\, ||\pmb{H}||_{\rm F}^2}_{\gamma_{\rm c}^{\rm L}}\leq\gamma_{\rm c}\leq \underbrace{\rho\, ||\pmb{H}||_{\rm F}^2}_{\gamma_{\rm c}^{\rm U}} .
\end{equation}
The distribution of the SNR $\gamma_{\rm c}$ in (\ref{eq:bounds_SNR2}) is not known, but we can bound its CDF by observing that, given two random variables $X$ and $Y$ such that $X\leq Y$ for any realizations, their CDFs are related as $F_X(x)\geq F_Y(x)$~\cite{Pa91}. Since the distribution of $||\pmb{H}||_{\rm F}^2$ is chi-square with $2MN$ degrees of freedom for i.i.d. Rayleigh fading, $\gamma_{\rm c}^{\rm U}=\rho||\pmb{H}||_{\rm F}^2$ and $\gamma_{\rm c}^{\rm L}=\gamma_{\rm c}^{\rm U}/\min\{M,N\}$ have the same distribution but for the mean value parameter.

To proceed, it is convenient to set up a normalization condition on the channel matrix $\pmb{H}$ by assuming its i.i.d. zero-mean entries have unit variance. This normalization does not limit the generality of the following discussion because all the considered SNRs scale with this variance accordingly. Under this normalization
$$
\mathbb{E}\left\{||\pmb{H}||^2_{\rm F}\right\} =M \sum_{i^1}^M\sum_{j=1}^N \mathbb{E}\left\{|h_{ij}|^2\right\} = MN .
$$
Using (\ref{eq:bounds_SNR2}), the average SNR at the input of the demodulator can be bounded as
$$
\underbrace{\rho\, \max\{M,N\}}_{\overline{\gamma}_{\rm c}^{\rm L}} \leq \overline{\gamma}_{\rm c} \leq \underbrace{\rho\, M\, N}_{\overline{\gamma}_{\rm c}^{\rm U}}
$$
where we used the fact that, given two random variables $X$ and $Y$ with $X\leq Y$ for any realizations, their means are related as $\mathbb{E}\{X\}\leq\mathbb{E}\{Y\}$~\cite{Pa91} and we have defined the means $\overline{\gamma}_{\rm c}^{\rm L}=\mathbb{E}\{\gamma_{\rm c}^{\rm L}\}$ and $\overline{\gamma}_{\rm c}^{\rm U}=\mathbb{E}\{\gamma_{\rm c}^{\rm U}\}$.

Based on the above setting, we can define lower and upper bounds on the CDF of the SNR $\gamma_{\rm c}$, denoted respectively as
$F_{\rm \ell}(\cdot)$ and $F_{\rm u}(\cdot)$, as chi-square with $2MN$ degrees of freedom and average per-branch SNRs $\overline{\gamma}^{\rm U}$ and $\overline{\gamma}^{\rm L}$, respectively defined as
\begin{eqnarray*}
\overline{\gamma}^{\rm U} &=& \rho\\
\overline{\gamma}^{\rm L} &=& \frac{\rho}{\min\{M,N\}} .
\end{eqnarray*}
These CDFs can be expressed by (\ref{eq:CDF_MRC}) with $M$ replaced by $MN$ and $\overline{\gamma}$ replaced by $\overline{\gamma}^{\rm U}$ and $\overline{\gamma}^{\rm L}$, respectively. Hence the outage probability can be bounded by these CDFs.

We can now bound the outage capacity of the massive MIMO diversity channel, using (\ref{eq:outage_definition}), as
$$
C_\varepsilon^{\rm L}(\overline{\gamma}_{\rm c}^{\rm L})\leq C_\varepsilon^{\rm MIMO}(\overline{\gamma}_{\rm c}) \leq C_\varepsilon^{\rm U}(\overline{\gamma}_{\rm c}^{\rm U})
$$
with
\begin{eqnarray*}
C_\varepsilon^{\rm U}(\overline{\gamma}_{\rm c}^{\rm U}) &=& \log_2\left(1+F_{\rm \ell}^{-1}(\varepsilon)\right) \\
C_\varepsilon^{\rm L}(\overline{\gamma}_{\rm c}^{\rm L}) &=& \log_2\left(1+F_{\rm u}^{-1}(\varepsilon)\right) .
\end{eqnarray*}
We can conclude that the outage capacity of the MIMO channel with full diversity is upper and lower bounded by that of the MRC SIMO channel with $MN$ antennas and per-branch average SNRs $\overline{\gamma}^{\rm U}$ and $\overline{\gamma}^{\rm L}$, respectively. This implies that the analysis in Section~\ref{subsec:outage_SIMO} for MRC allows to derive the above bounds, provided the correct diversity order and SNRs are considered.

These bounds may be reasonably tight if $\min\{M,N\}$ is small, i.e., if a large number of antennas is used at one side only, either transmitter or receiver. However, for doubly massive systems in which both $M$ and $N$ are large, $\min\{M,N\}$ is large and the bounds become loose. To derive a capacity benchmark useful for the the doubly massive MIMO diversity channel, we may resort to the asymptotic results in~\cite{Ed88} which show that the largest squared singular value if $M$ and $N$ are both large, but $y=M/N$ is fixed, approaches
$$
\sigma^2_{\rm max} \rightarrow 2(1+\sqrt{y})^2 N .
$$
This implies that
\begin{equation} \label{eq:Edelman}
\overline{\gamma}_{\rm c} \rightarrow \rho\,  2(1+\sqrt{y})^2 N 
\end{equation}
and, therefore, the outage capacity tends to that of MRC with $2(1+\sqrt{y})^2 N$ antennas.

\section{Numerical Results}\label{sec:results}
We now present numerical results on the analytical framework presented in Section~\ref{sec:outage_capacity}. We consider two representative values of high and low outage probabilities, namely $\varepsilon=10^{-1}$ and $\varepsilon=10^{-3}$, respectively. In Section~\ref{subsec:results_high_snr}, we focus on the high-SNR regime, where the capacity gap is a representative performance indicator. In Section~\ref{subsec:results_low_snr}, we discuss the low-SNR regime where the capacity ratio is of interest.

\subsection{High-SNR Analysis} \label{subsec:results_high_snr}
\begin{figure}
\centering
\includegraphics[width=0.48\textwidth]{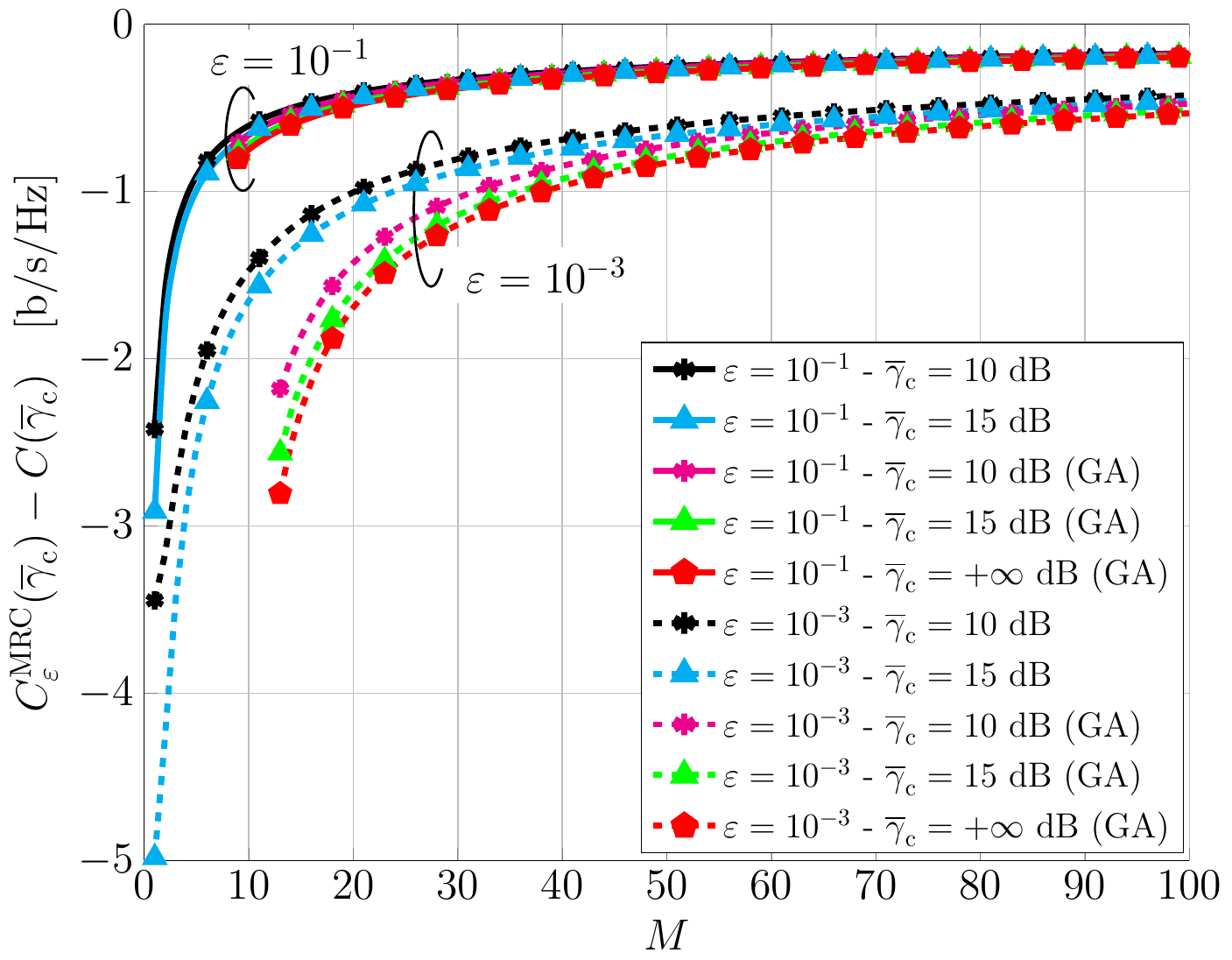}
\caption{Capacity gap with respect to the benchmark $C(\overline{\gamma}_{\rm c})$, as a function of $M$, for the SIMO channel with MRC and various (large) values of SNR $\overline{\gamma}_{\rm c}$ and outage probability $\varepsilon$. The results obtained by numerical inversion of the SNR CDF are compared with the GA.}
\label{fig:outage_capacity_MRC}
\end{figure}

In Fig.~\ref{fig:outage_capacity_MRC}, the capacity gap with respect to the AWGN channel with SNR $\overline{\gamma}_{\rm c}$ is shown, as a function of $M$, for the SIMO channel with MRC and various (large) values of SNR $\overline{\gamma}_{\rm c}$ and outage probability $\varepsilon$. The results obtained by numerical inversion of the SNR CDF are compared with the GA.
All theoretical predictions can be confirmed---in particular, the larger the number of antennas, the closer to zero the capacity gap. It is worth noting the slow convergence to this value as $M^{-1/2}$. As expected, the smaller the outage probability, the worse the performance. Finally, note that for large values of $M$, the GA well predicts the system performance.

\begin{figure}
\centering
\includegraphics[width=0.48\textwidth]{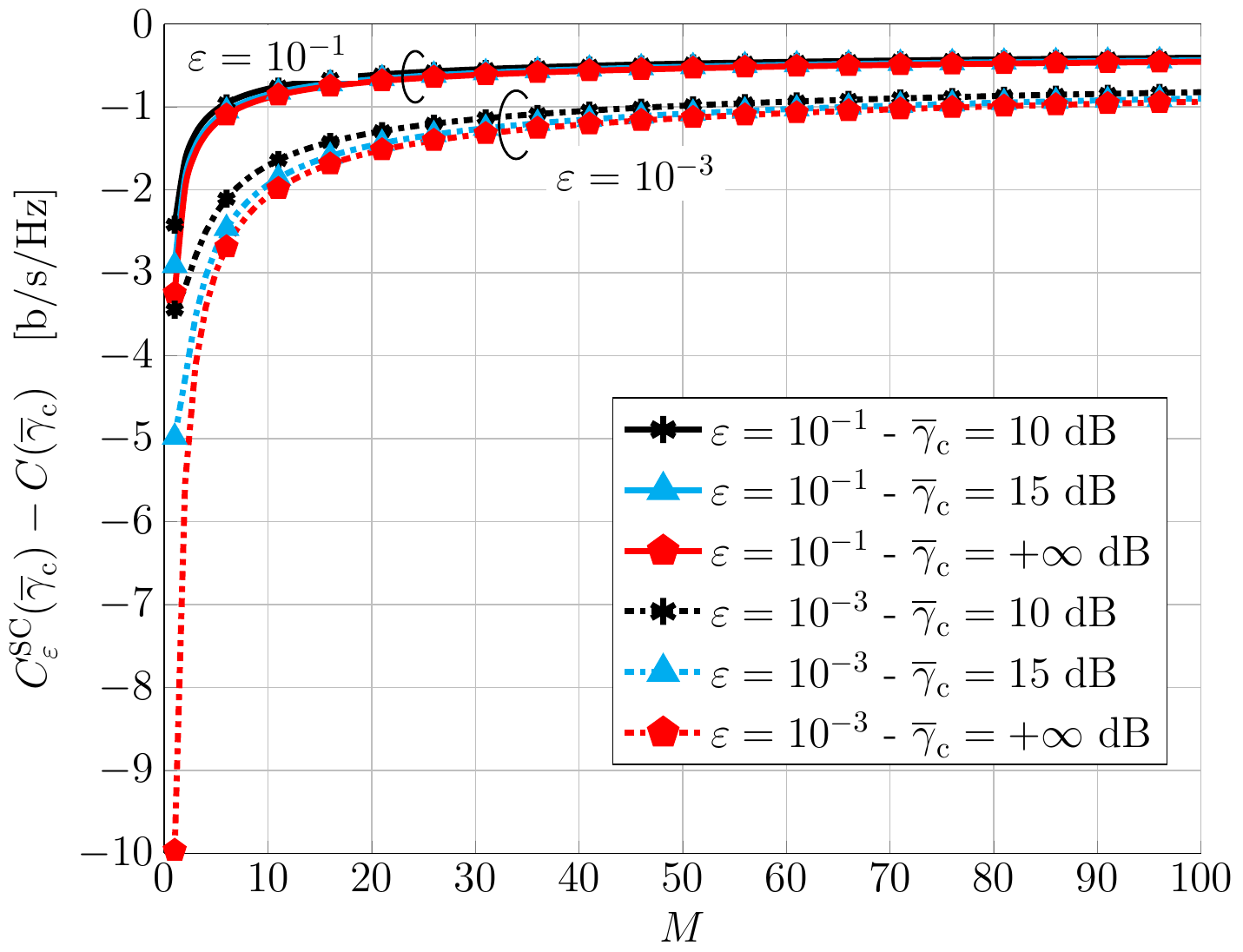}
\caption{Capacity gap with respect to the benchmark $C(\overline{\gamma}_{\rm c})$, as a function of $M$, for the SIMO channel with SC and various (large) values of SNR $\overline{\gamma}_{\rm c}$ and outage probability $\varepsilon$.}
\label{fig:outage_capacity_SC}
\end{figure}

In Fig.~\ref{fig:outage_capacity_SC}, the capacity gap with respect to the AWGN channel with SNR $\overline{\gamma}_{\rm c}$, as a function of $M$, is shown for the SIMO channel with SC and various (large) values of SNR $\overline{\gamma}_{\rm c}$ and outage probability $\varepsilon$.
Similar conclusions as in MRC can be drawn for SC. The curves for increasing value of SNR approach the benchmark $C(\overline{\gamma}_{\rm c})$ as predicted by (\ref{eq:outage_SC_Lfix}). As $M$ increases, a slow growth can be observed. This means that the benchmark can only be approached for realistic values of $M$ except for a gap. As an example, considering $\varepsilon=10^{-3}$, $C_\varepsilon^{\rm SC}-C(\overline{\gamma}_{\rm c})\simeq-0.94$~b/s/Hz for $M=100$. Increasing $M$ from 100 to 1000 leads to a reduction in the capacity difference to approximately -0.59. Further increasing $M$ to 10000 leads to a gap of approximately \mbox{-0.43}. Recall that $M=10000$ can still be considered realistic, as per the example discussed in Section~\ref{sec:system-model}.

The results in Figs.~\ref{fig:outage_capacity_MRC} and~\ref{fig:outage_capacity_SC} are also valid for the MISO channel with MRT and ST with $M$ transmit antennas, respectively.

\begin{figure}
\centering
\includegraphics[width=0.48\textwidth]{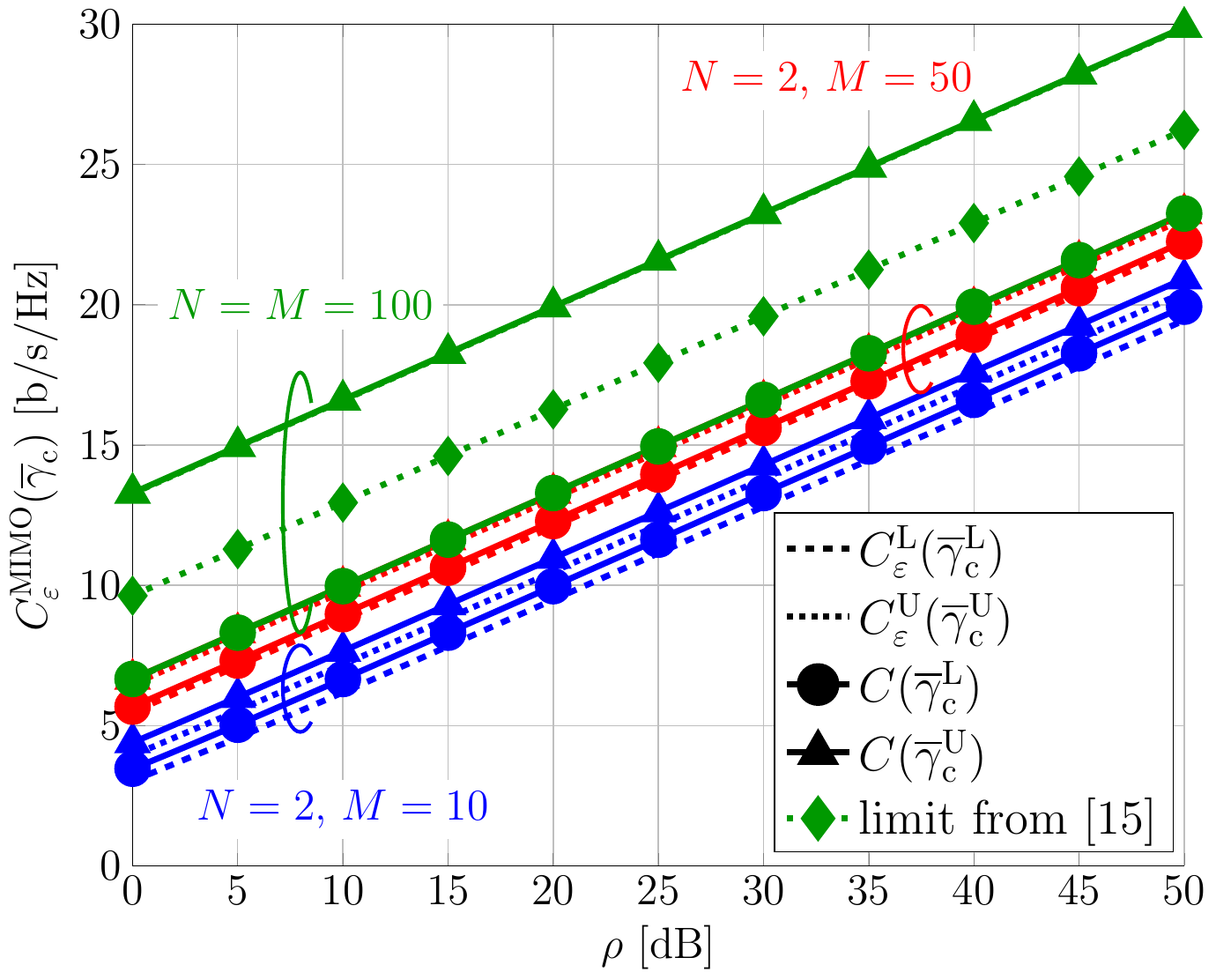}
\caption{Outage capacity, as a function of SNR, for the massive MIMO diversity channel with $\varepsilon=10^{-1}$ and various values of $N$ and $M$. Upper and lower bounds, as well as benchmarks are shown.}
\label{fig:outage_capacity_MIMO}
\end{figure}

Fig.~\ref{fig:outage_capacity_MIMO} shows the outage capacity, as a function of the SNR, for the massive MIMO diversity channel with $\varepsilon=10^{-1}$ and various values of $N$ and $M$.\footnote{Similar considerations hold for other values of the outage probability.} Upper and lower bounds, as well as benchmarks are shown.
As one can see, for all the considered scenarios, upper and lower bounds are close to the benchmarks $C(\overline{\gamma}_{\rm c}^{\rm U})$ and $C(\overline{\gamma}_{\rm c}^{\rm L})$. Moreover, for relatively small values of $M$, the bounds are close to each other and the performance is well predicted. On the other hand, in the doubly massive case $N=M=100$, the performance is well predicted by the asymptotic result from~\cite{Ed88}.

\subsection{Low-SNR Analysis} \label{subsec:results_low_snr}
\begin{figure}
\centering
\includegraphics[width=0.48\textwidth]{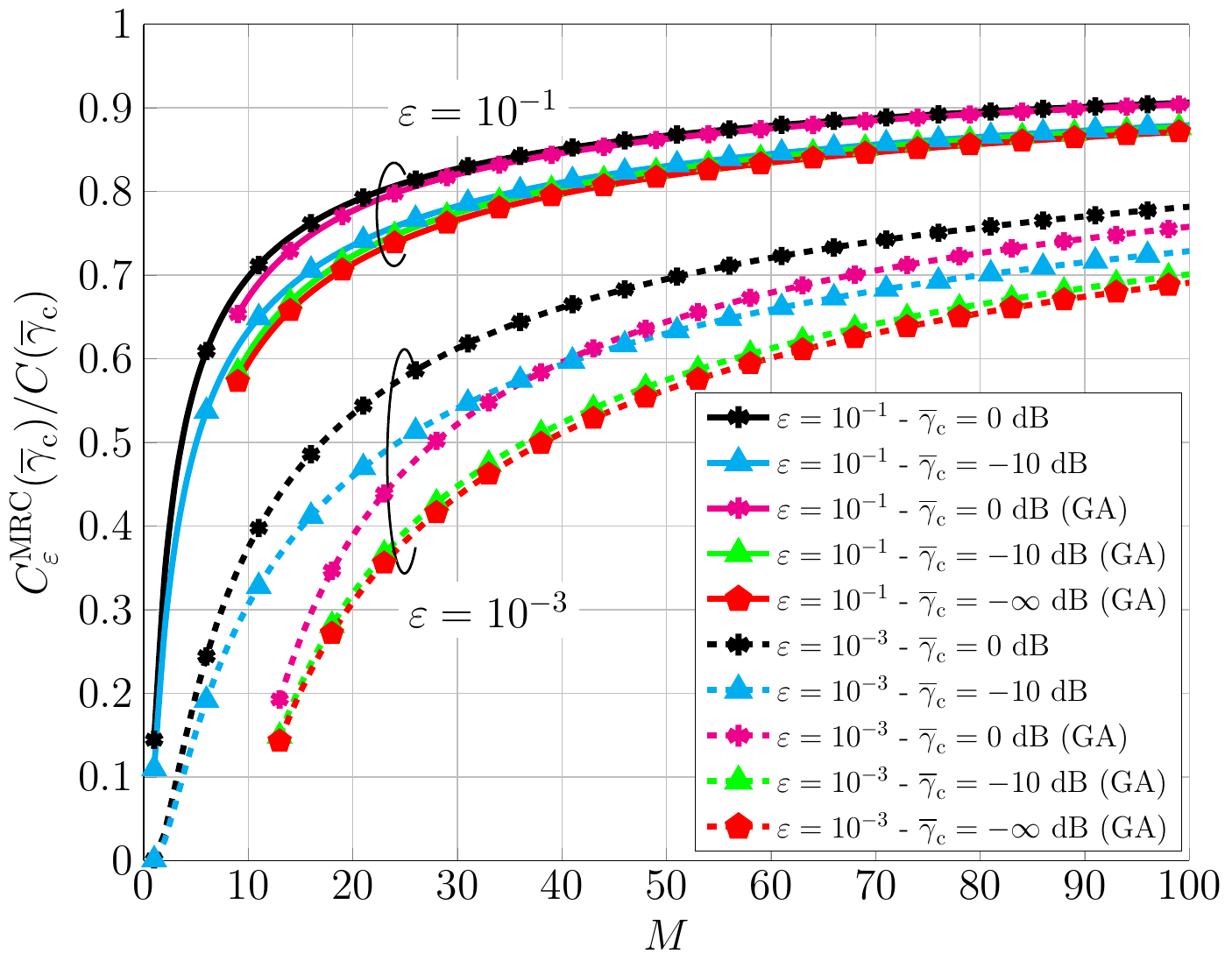}
\caption{Capacity ratio, as a function of $M$, for the SIMO channel with MRC and various (low) values of SNR $\overline{\gamma}_{\rm c}$ and outage probability $\varepsilon$. The results obtained by numerical inversion of the SNR CDF are compared with the GA.}
\label{fig:outage_capacity_MRC_low}
\end{figure}

In Fig.~\ref{fig:outage_capacity_MRC_low}, the capacity ratio is shown, as a function of $M$, for the SIMO channel with MRC and various (low) values of SNR $\overline{\gamma}_{\rm c}$ and outage probability $\varepsilon$. The results obtained by numerical inversion of the SNR CDF are compared with the GA.
Similar conclusions as for the capacity gap in Fig.~\ref{fig:outage_capacity_MRC} can be drawn. One should observe that for large values of $M$, the GA well predicts the system performance in this case as well. In particular, at low SNR the GA approaches 1 as $M^{-1/2}$ for realistic values of $M$. Note that compared with the reference value 1, i.e., $C_\varepsilon^{\rm MRC}=C(\overline{\gamma}_{\rm c})$, the outage capacity is only a fraction and the loss is more evident with respect to the high-SNR case shown in Fig.~\ref{fig:outage_capacity_MRC}.

\begin{figure}
\centering
\includegraphics[width=0.48\textwidth]{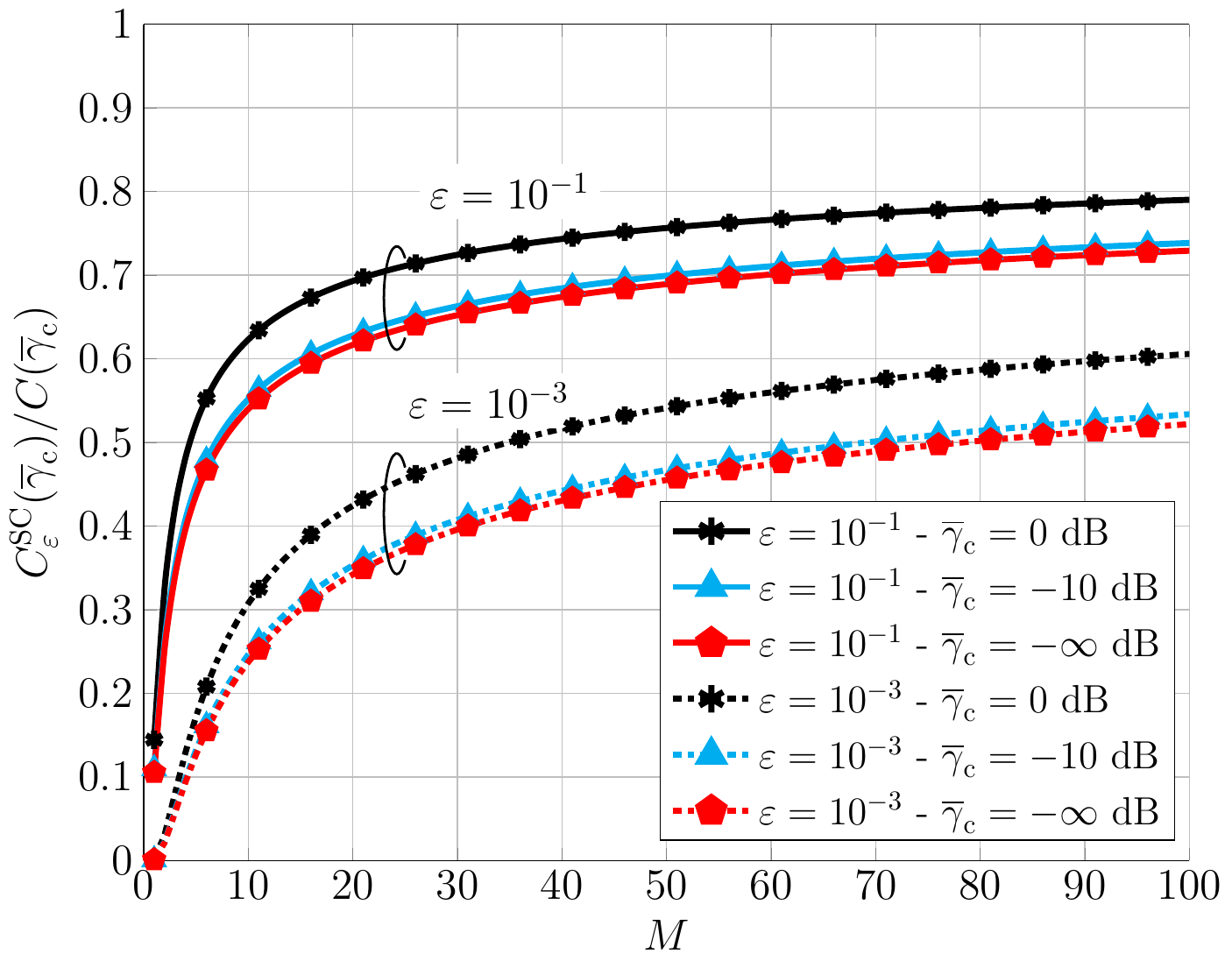}
\caption{Capacity ratio, as a function of $M$, for the SIMO channel with SC and various (low) values of SNR $\overline{\gamma}_{\rm c}$ and outage probability $\varepsilon$.}
\label{fig:outage_capacity_SC_low}
\end{figure}

In Fig.~\ref{fig:outage_capacity_SC_low}, the capacity ratio is shown, as a function of $M$, for the SIMO channel with SC and various (low) values of SNR $\overline{\gamma}_{\rm c}$ and outage probability $\varepsilon$.
As for the capacity gap in Fig.~\ref{fig:outage_capacity_SC}, as $M$ increases very slow (logarithmic) growth can be observed and the benchmark cannot be approached in practice. Moreover, the fraction of the reference value 1, i.e., $C_\varepsilon^{\rm SC}=C(\overline{\gamma}_{\rm c})$, is lower than for MRC, meaning that SC is less effective in the low-SNR regime.

The results in Figs.~\ref{fig:outage_capacity_MRC_low} and~\ref{fig:outage_capacity_SC_low} are also valid for the MISO channel with MRT and ST with $M$ transmit antennas, respectively.

\begin{figure}
\centering
\includegraphics[width=0.48\textwidth]{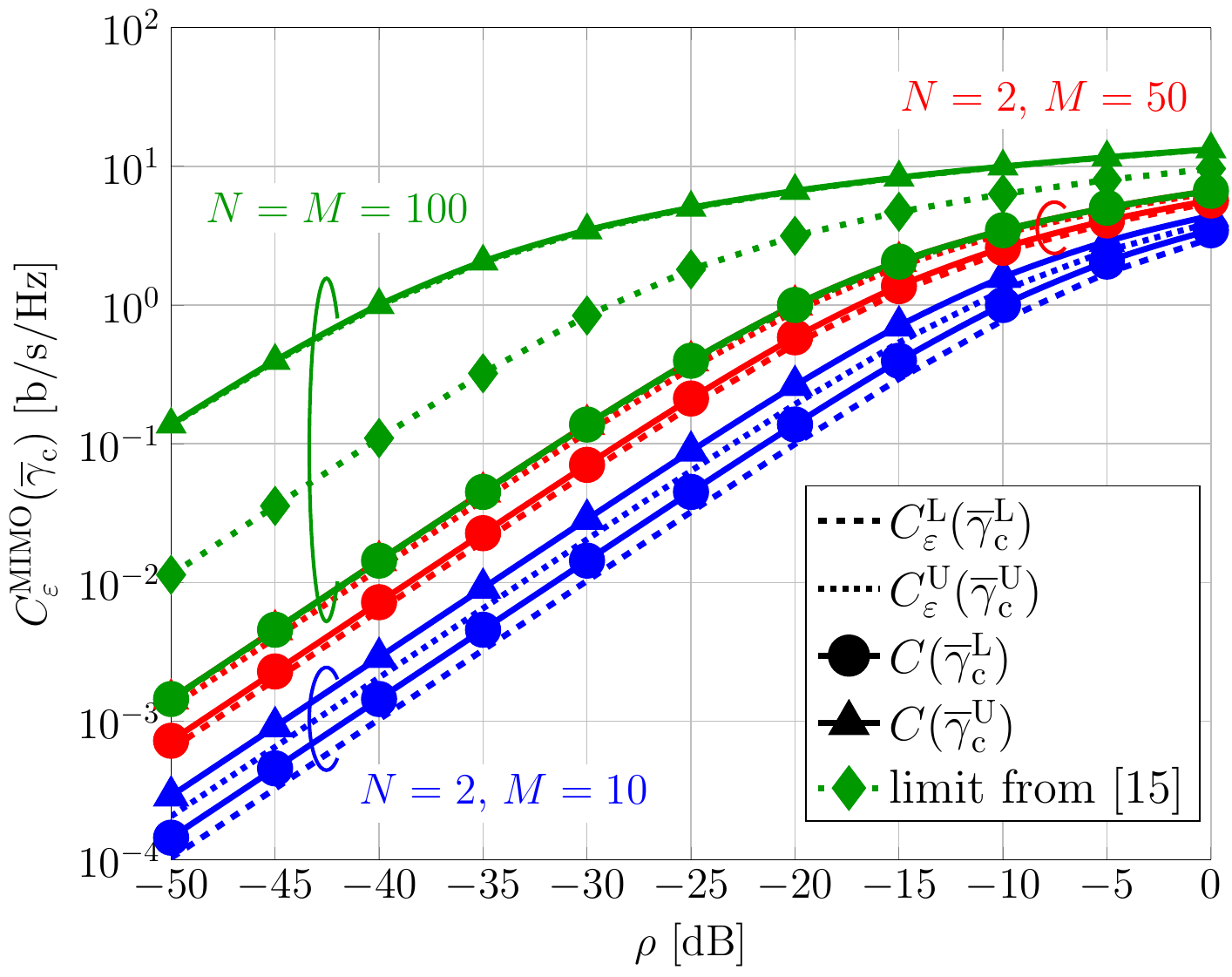}
\caption{Outage capacity, as a function of SNR, for the massive MIMO diversity channel with $\varepsilon=10^{-1}$ and various values of $N$ and $M$. Upper and lower bounds, as well as benchmarks are shown.}
\label{fig:outage_capacity_MIMO_low}
\end{figure}

In Fig.~\ref{fig:outage_capacity_MIMO_low}, the outage capacity in the low SNR regime is shown for the massive MIMO diversity channel with $\varepsilon=10^{-1}$ and various values of $N$ and $M$. Upper and lower bounds, as well as benchmarks are shown.
Similar considerations to those in Fig.~\ref{fig:outage_capacity_MIMO} for the high-SNR regime are valid in this case for the low-SNR regime as well.

\section{Concluding Remarks} \label{sec:concl}
In this paper, we presented an analysis of the outage capacity for the massive MIMO diversity channel subject to i.i.d. Rayleigh fading. We started from a SIMO channel, providing a numerical solution and a GA for MRC, whereas an exact analysis was presented for SC. The analysis was shown to be valid for the MISO channel with MRT and ST. The analysis was then extended to the massive MIMO diversity channel, showing that its outage capacity with full diversity is upper and lower bounded by that of an MRC system with a proper number of antennas. Our results show that, if the number of antennas is sufficiently large, the capacity of each considered diversity channel approaches that of suitable reference AWGN channels with properly defined SNRs under realistic number of antennas. Finally, proper bounds and an asymptotic benchmark are provided for the massive MIMO case.

\bibliographystyle{IEEEtran}
\bibliography{references}

\begin{thebibliography}{10}
\providecommand{\url}[1]{#1}
\csname url@samestyle\endcsname
\providecommand{\newblock}{\relax}
\providecommand{\bibinfo}[2]{#2}
\providecommand{\BIBentrySTDinterwordspacing}{\spaceskip=0pt\relax}
\providecommand{\BIBentryALTinterwordstretchfactor}{4}
\providecommand{\BIBentryALTinterwordspacing}{\spaceskip=\fontdimen2\font plus
\BIBentryALTinterwordstretchfactor\fontdimen3\font minus
  \fontdimen4\font\relax}
\providecommand{\BIBforeignlanguage}[2]{{%
\expandafter\ifx\csname l@#1\endcsname\relax
\typeout{** WARNING: IEEEtran.bst: No hyphenation pattern has been}%
\typeout{** loaded for the language `#1'. Using the pattern for}%
\typeout{** the default language instead.}%
\else
\language=\csname l@#1\endcsname
\fi
#2}}
\providecommand{\BIBdecl}{\relax}
\BIBdecl

\bibitem{TsVi12}
D.~Tse and P.~Viswanathan, \emph{Fundamentals of Wireless Communication}.\hskip
  1em plus 0.5em minus 0.4em\relax New York, NY, USA: Cambridge University
  Press, June 2012, {DOI: 10.1017/CBO9780511807213}.

\bibitem{BaShKa20}
O.~S. {Badarneh}, M.~K. {Shawaqfeh}, and M.~{Kadoch}, ``Performance analysis of
  mobile {IoT} networks over composite fading channels,'' in \emph{Proc. Int.
  Wireless Communications and Mobile Computing {\rm (IWCMC)}}, Limassol, Cyprus
  (held as virtual), June 2020, pp. 1234--1239, {DOI:
  10.1109/IWCMC48107.2020.9148477}.

\bibitem{ZhBjMaNgYaLo20}
J.~{Zhang}, E.~{Bj\"ornson}, M.~{Matthaiou}, D.~W.~K. {Ng}, H.~{Yang}, and
  D.~J. {Love}, ``Prospective multiple antenna technologies for beyond {5G},''
  \emph{{IEEE} J. Select. Areas Commun.}, vol.~38, no.~8, pp. 1637--1660,
  August 2020, {DOI: 10.1109/JSAC.2020.3000826}.

\bibitem{KhViYu21}
S.~R. {Khosravirad}, H.~{Viswanathan}, and W.~{Yub}, ``Exploiting diversity for
  ultra-reliable and low-latency wireless control,'' \emph{{IEEE} Trans.
  Wireless Commun.}, vol.~20, no.~1, pp. 316--331, January 2021, {DOI:
  10.1109/TWC.2020.3024741}.

\bibitem{PoStNiCaAnTrBa19}
P.~{Popovski}, A.~{Stefanovic}, J.~J. {Nielsen}, E.~{de Carvalho},
  M.~{Angjelichinoski}, K.~F. {Trillingsgaard}, and A.~{Bana}, ``Wireless
  access in ultra-reliable low-latency communication {(URLLC)},'' \emph{{IEEE}
  Trans. Commun.}, vol.~67, no.~8, pp. 5783--5801, August 2019, {DOI:
  10.1109/TCOMM.2019.2914652}.

\bibitem{JoWaErHe15}
N.~A. {Johansson}, Y.~.~E. {Wang}, E.~{Eriksson}, and M.~{Hessler}, ``Radio
  access for ultra-reliable and low-latency {5G} communications,'' in
  \emph{Proc.~IEEE Intern. Conf. on~Commun. {\rm (ICC)}}, London, UK, June
  2015, pp. 1184--1189, {DOI: 10.1109/ICCW.2015.7247338}.

\bibitem{DiQuLiCh21}
J.~{Ding}, D.~{Qu}, P.~{Liu}, and J.~{Choi}, ``Machine learning enabled
  preamble collision resolution in distributed massive {MIMO},'' \emph{{IEEE}
  Trans. Commun.}, 2021, {to appear. DOI: 10.1109/TCOMM.2021.3051202}.

\bibitem{LiLvZhWaWaYo20}
J.~{Li}, Q.~{Lv}, P.~{Zhu}, D.~{Wang}, J.~{Wang}, and X.~{You},
  ``Network-assisted full-duplex distributed massive {MIMO} systems with
  beamforming training based {CSI} estimation,'' \emph{{IEEE} Trans. Wireless
  Commun.}, 2020, {to appear. DOI: 10.1109/TWC.2020.3040044}.

\bibitem{BuDa18}
S.~{Buzzi} and C.~{D'Andrea}, ``Energy efficiency and asymptotic performance
  evaluation of beamforming structures in doubly massive {MIMO mmWave}
  systems,'' \emph{IEEE Trans. Green Commun. and Networking}, vol.~2, no.~2,
  pp. 385--396, June 2018, {DOI: 10.1109/TGCN.2018.2800537}.

\bibitem{Lo99}
T.~K.~Y. Lo, ``Maximum ratio transmission,'' \emph{{IEEE} Trans. Commun.},
  vol.~47, no.~10, pp. 1458--1461, October 1999, {DOI: 10.1109/26.795811}.

\bibitem{Br03}
D.~G. {Brennan}, ``Linear diversity combining techniques,'' \emph{Proc.
  {IEEE}}, vol.~91, no.~2, pp. 331--356, February 2003, {DOI:
  10.1109/JPROC.2002.808163}.

\bibitem{Ha10}
J.~Havil, \emph{Gamma: Exploring Euler's Constant}.\hskip 1em plus 0.5em minus
  0.4em\relax Princeton, NJ, USA: Princeton University Press, 2010, {DOI:
  10.1515/9781400832538}.

\bibitem{PaNaGo03}
A.~Paulraj, R.~Nabar, and D.~Gore, \emph{Introduction to Space-Time Wireless
  Communications}.\hskip 1em plus 0.5em minus 0.4em\relax Cambridge, UK:
  Cambridge University Press, 2003.

\bibitem{Pa91}
A.~Papoulis, \emph{{Probability, Random Variables and Stochastic
  Processes}}.\hskip 1em plus 0.5em minus 0.4em\relax New York, NY, USA:
  McGraw-Hill, 1991.

\bibitem{Ed88}
A.~Edelman, ``Eigenvalues and condition numbers of random matrices,''
  \emph{SIAM J. Matrix Analysis and Appl.}, vol.~9, no.~4, pp. 543--560, 1988,
  {DOI: 10.1137/0609045}.

\end{thebibliography}

\end{document}